\documentclass[a4paper]{article}

\usepackage{INTERSPEECH2018}

\title{AVA-Speech: A Densely Labeled Dataset of Speech Activity in Movies}
\name{Sourish Chaudhuri, Joseph Roth, Daniel P. W. Ellis, Andrew Gallagher, Liat Kaver, Radhika Marvin, Caroline Pantofaru, Nathan Reale, Loretta Guarino Reid, Kevin Wilson, Zhonghua Xi}
\address{Google Inc.}
\email{\{sourc, josephroth, dpwe, agallagher, lkaver, radahika, cpantofaru, nreale, lorettaguarino, kwwilson, zxi\}@google.com}

\begin{document}

\maketitle
\begin{abstract}
Speech activity detection (or endpointing) is an important processing step for applications such as speech recognition, language identification and speaker diarization. Both audio- and vision-based approaches have been used for this task in various settings, often tailored toward end applications. However, much of the prior work reports results in synthetic settings, on task-specific datasets, or on datasets that are not openly available. This makes it difficult to compare approaches and understand their strengths and weaknesses. In this paper, we describe a new dataset which we will release publicly containing densely labeled speech activity in YouTube videos\footnotemark, with the goal of creating a shared, available dataset for this task. The labels in the dataset annotate three different speech activity conditions: clean speech, speech co-occurring with music, and speech co-occurring with noise, which enable analysis of model performance in more challenging conditions based on the presence of overlapping noise. We report benchmark performance numbers on AVA-Speech using off-the-shelf, state-of-the-art audio and vision models that serve as a baseline to facilitate future research.
\end{abstract}
\footnotetext{The speech activity labels are available on the AVA website at http://research.google.com/ava.}

\section{Introduction}
\label{sec:intro}
Speech activity detection, also called ``endpointing'' has been an essential component in processing pipelines for speech recognition, language identification, and speaker diarization, and has grown increasingly important with the growth of online media and voice-based interfaces. Approaches proposed for this task include signal-based and feature extraction-based analysis \cite{Che99, Shi00, Sad13, Woo00}, as well as machine learning \cite{Shi10, Ng12, Gha15}, with neural network-based approaches growing increasingly popular \cite{Rya13, Jan17, Maa17, Cha17}. While speech detection has traditionally been an audio task, many application domains such as web videos have associated video, and visual classification approaches have sought to improve over audio-only approaches in noisy environments \cite{Pet09, Dov15, Buc16, Tao17}. 

Depending on the specific application that speech activity detection is used for, developers make varying choices of the appropriate parameters that determine the trade-off between false alarms and missed detections for speech. In this context, a key limitation in the literature on this topic is the absence of a standard benchmark dataset for direct comparisons between models. Given the context of the variety of approaches in the literature, an ideal dataset would have the following characteristics: 

\begin{itemize}
\item Should contain video, so audio and visual (and audio-visual) methods can be compared on the same data. 
\item Should be densely labeled so that each instant corresponds to a label.
\item Should contain real data corresponding to at least one of the typical use cases for such systems, {\it e.g.} natural conversational setting, short utterances typically used in navigating intelligent voice interfaces, etc.
\item Should have a natural mix of background noise conditions, as opposed to synthetic, controlled addition of noise for evaluation. 
\end{itemize}

It is likely that the difficulty of developing a relevant dataset that satisfies all the conditions above has led to the community not having converged to a clear choice for a benchmark. Our goal in this work is to develop such a benchmark dataset, which we call AVA-Speech\footnotemark.
\footnotetext{The `AVA' in the name refers to the corpus used as the source of videos for labeling, which we discuss in Section \ref{sec:labeling}.}
Our hope is that it will be not only a benchmark for speech detection in the near-term but a dataset that can be actively developed to add labels for tasks beyond speech activity detection including joint audio-visual modeling due to the availability of video.

Prior stand-alone speech detection work has taken a variety of approaches to reporting speech detection model performance, including datasets containing individual utterances (e.g., from TIMIT) \cite{Che99, Shi00, Tao17}, with noise added to them \cite{Pet09, Dov15}, and datasets that are not easily publicly available \cite{Rya13, Maa17, Cha17}; most of these are also audio-only datasets. Of the few publicly available datasets, a popular choice for speech activity detection model evaluation (used in \cite{Gha15}), the QUT-NOISE-TIMIT corpus \cite{Dea10} contains an artificial sequential combination of individual utterances, as well as added noise recorded from 5 scenarios.

Literature involving work around conversational datasets (usually, as part of speaker diarization \cite{Ang06, Mcc05, Gal09, Zel12}) have used more realistic data from meeting and broadcast news (BN) contexts. Meeting datasets are typically more spontaneous in their content but are limited along the axes described above: video data is rare, context and participant sets are small due to the logistical difficulties involved, and noise conditions are often solely room reverberations. BN is a better fit in terms of diversity of speakers but not for diversity of noise conditions, since the majority of the data are in-studio conversations. Datasets from movies and TV shows\textemdash ETAPE corpus (7 hours from 15 TV shows) \cite{Gra12}, REPERE corpus (3 hours from 28 TV shows) \cite{Gir12}, 4 Hollywood movies in \cite{Leh15} \textemdash comes closest to satisfying the characteristics described above, and are fairly close in the style of content to the labeled dataset that we contribute in this work. However, AVA-Speech is 5${\times}$ larger than the largest of the TV/movies datasets and contains data from 190 movies, and should lead to a wider diversity of contexts. It explicitly annotates when speech activity co-occurs with background sounds to call out the challenging cases for speech detectors. Unlike datasets that were purposely recorded for a specific task, we had no influence over recording conditions, video production or narrative structures used. As such, we believe this dataset should serve well as a general evaluation benchmark for analysis of open domain media content on the web. Through the rest of this paper, we describe our work in developing AVA-Speech, which will be released publicly. Specifically, we discuss:

\begin{itemize}
\item The choice of the (videos in the) dataset, for which we will provide YouTube video IDs (15 minute clips from 185 movies for ~45 hours total) along with manually annotated dense labels indicating the presence or absence of speech activity. This satisfies the characteristics 1, 2 and 4 above and satisfies 3 to the extent possible by videos from the movies domain.

\item The labeling explicitly annotates segments containing active speech as one of three classes\textemdash clean speech, speech co-occurring with music or speech co-occurring with noise \textemdash and the ones that don't contain speech. We discuss our choice of labels and the labeling instructions provided to human raters.

\item Since we had no control over the process of production of the videos, we do not have ground-truth speech to noise ratios, and cannot easily characterize the level of background noise that could be challenging for an audio-based detector. We do, however, provide an estimate of the speech to noise ratio using a neural network-based speech enhancement model.

\item We present audio-only and vision-only performance metrics on AVA-Speech using state-of-the-art (but off-the-shelf) audio and vision systems (i.e., they were not optimized for AVA-Speech) that can serve as baselines for future comparisons.

\end{itemize}

\begin{figure}[t]
  \centering
  \includegraphics[width=\linewidth]{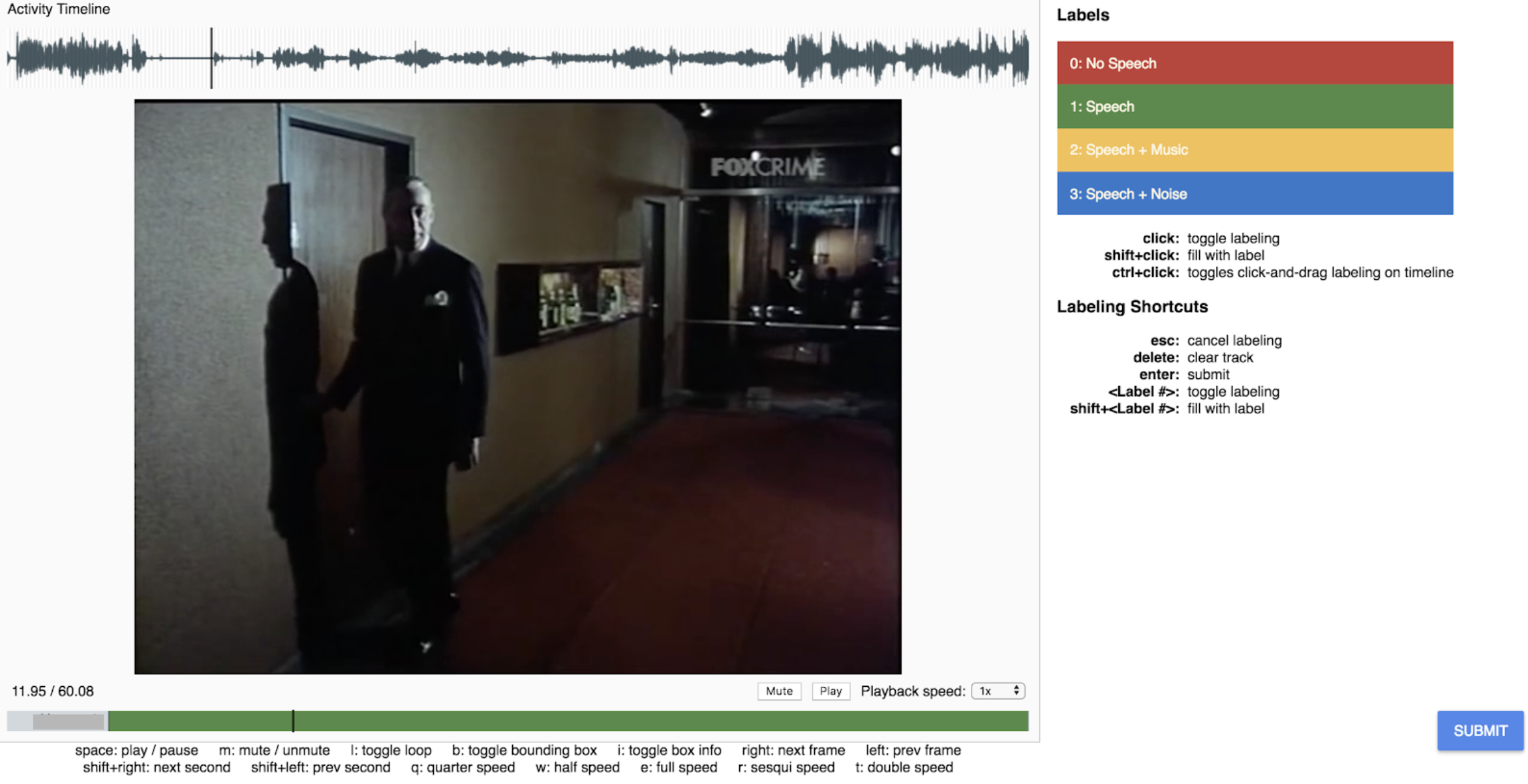}
  \caption{Rating interface with the set of labels and labeling shortcuts shown on the right of the video player, and playback shortcuts below the player.}
  \label{fig:rating_ui}
\end{figure}

\begin{figure}[t]
  \centering
  \includegraphics[width=\linewidth]{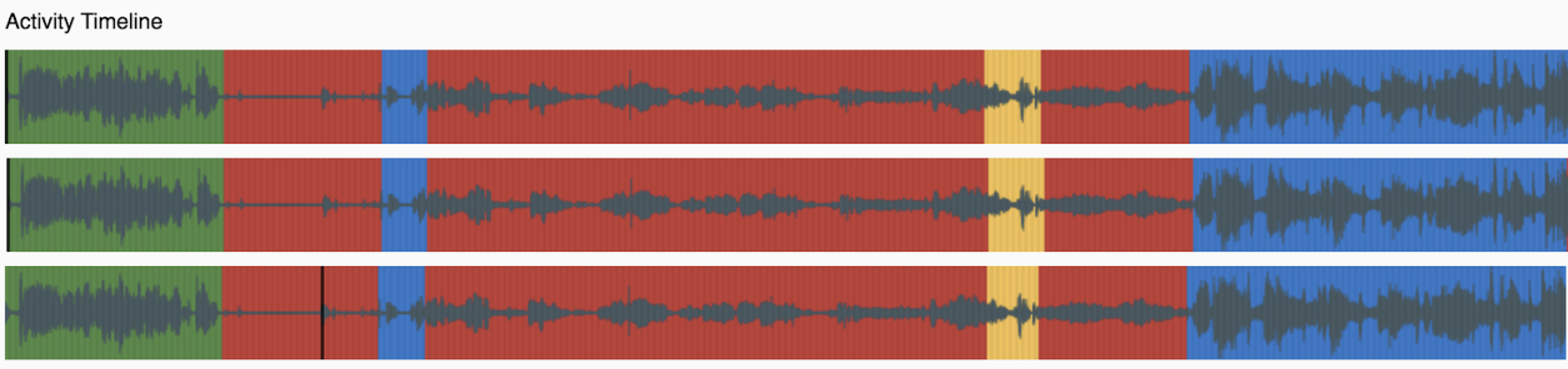}
  \caption{An example of the labeled activity timeline from 3 labelers for the clip in Figure \ref{fig:rating_ui}. In general, inter-labeler agreement is high for the speech segment labeling task.}
  \label{fig:rated_segment}
\end{figure}



The rest of this paper is organized as follows: Section \ref{sec:labeling} describes the choice of dataset and videos, choices of labels, and the human labeling process. Section \ref{sec:datastats} discusses various statistics relevant to the dataset including estimation of the SNR across the different label classes. Section \ref{sec:models} presents audio and visual models and their performance on this dataset, and we conclude with a discussion in Section \ref{sec:conclusion}.

\section{Dataset and Labels}
\label{sec:labeling}

The video clips in this dataset are from the AVA dataset \cite{Gu18} v1.0 (hence the name, AVA-Speech). The AVA dataset is sourced from 192 movies on YouTube, and contain continuous segments between minutes 15-30 of each movie. Please see Section 3 of \cite{Gu18} for details of the video selection process. While movies are not a perfect representation of {\it in-the-wild} broadcast media content, we chose to use these videos for the following reasons.

First, movies have diversity in acoustic and visual scenes, speakers and speaker demographics. The movies in the AVA dataset were chosen from international film industries, and include movies in multiple languages and movies with dubbed audio, similar to the content we would expect to see in unconstrained media on the Internet. 

Second, the audio speech labels described in this paper complement the visual action recognition annotations that already exist for AVA. Having a dataset with both audio and visual labels allows the two communities to work closer together on shared problems, and promotes the development and testing of audio-visual multimodal models. Going forward, multimodal model exploration should go beyond speech activity detection, for example the tasks in \cite{Mro15, Chu16, Hoo18}. The data annotations will also facilitate deeper and better understanding of how audio speech correlates with visual content.

Finally, movies present a ready opportunity to serve as a potential dataset for applications such as speech recognition or diarization, due to the presence of a structured narrative with conversations in different contexts - room and noise conditions, and varying groups of participants and scene structures.

We also note that this dataset differs in significant ways from a pair of recently released YouTube-based datasets. The original AVA labels released in \cite{Gu18} also contain labels for the {\it talk to} activity that can be considered analogous to spoken activity; however, the labelers in \cite{Gu18} only had access to the visual stream, and thus the {\it talk to} label corresponds to visual speech - when the speaker's face can be seen, regardless of whether they can be heard - rather than audible speech, and is only annotated at 1 frame per second. AudioSet \cite{gemmeke2017audio} also contains {\it speech} labels, but differs from AVA-Speech in two significant ways: (1) It is labeled at the 10-second clip level, without more accurate temporal localization within the clip; (2) It focuses on a wide variety of audio events beyond speech, and the speech clips were not chosen with potential applicability to downstream tasks, such as speech recognition or diarization in mind. 

In our dataset, the labels are: NoSpeech, CleanSpeech, Speech+Music and Speech+Noise. We broke out the speech activity category into 3  classes since the presence of background sounds negatively affects audio-based detection models. Music is separated from all other sounds since it is a particularly difficult distractor and often co-occurs with speech; e.g., in video of a party, a movie musical, or as a narrative tool. 

\subsection{Human Labeling Interface and Instructions} 

Figure \ref{fig:rating_ui} shows the labeling interface. Raters are initially presented with an empty activity timeline above the player, showing the audio waveform. They select a label (from the right) and its start- and end-points on the timeline, and proceed to label the entire timeline. Labels are mutually exclusive. Examples are shown below the rating interface in Figure \ref{fig:rated_segment}.

In the labeling guidance, there are a few aspects worth highlighting. Labelers were explicitly asked to annotate all audible speech. This directly disadvantages visual-only speech detection models on this dataset, since speech may have been heard but not seen, and this bias is opposite to the AVA v1.0 visual labels. They were asked to mark activities with spoken communication intent as one of the speech categories, including garbled speech, unintelligible speech, foreign language speech, filler words such as ``um'',``ah'', etc. that were part of spoken communication, singing, and speech from electronic devices. Examples of audio that should be labeled as NoSpeech included sighs, coughs, grunts, and laughs. 

Finally, the instructions provided included guidance for differentiating between the three speech subclasses. Speech+Music indicates the presence of music as the only other sound alongside active speech, and a capella, rap music, and music with lyrics all belong to this class. Speech+Noise indicates the simultaneous presence of other non-music sounds, perhaps also including music. Labelers were also provided a label precedence for confusing situations: Speech+Noise $>$ Speech+Music $>$ CleanSpeech $>$ NoSpeech.

\section{Dataset Statistics}
\label{sec:datastats}
For human annotation, the 15-minute movie segments were subdivided into 1-minute clips, and each clip was annotated by 3 human labelers. The 3 ratings were merged at the (video) frame level using a majority vote. To compute inter-labeler agreement, we used Fleiss' $kappa$ \cite{Fle71} as each question required 3 labelers, and the pool comprised 20 labelers. The agreement between labelers was well above chance, with a $kappa$ value of 0.74 indicating ``substantial'' agreement (0.0 is no agreement above chance, and 1.0 is perfect agreement) measured using each video frame as a data instance and the four labels as the available categorical ratings. Figure \ref{fig:rated_segment} shows an example of the labels from 3 labelers on a single 1 minute clip. We note that the AVA dataset v1.0 consisted of 192 videos but 7 of those videos are no longer available on YouTube and are not included here.

First, we look at aggregate statistics of the labels in AVA-Speech (this dataset) in Table \ref{tab:data-stats}. The dataset is roughly evenly split between speech and no-speech, by time and by number of segments, unlike other datasets that are speech-heavy. Also note that the data are not biased towards clean speech, instead there is ~2.5$\times$ as much speech co-occurring with background noise. Both of these attributes make the dataset generally interesting for downstream applications that use speech activity detection as a component, such as diarization.

The original release of the AVA dataset v1.0 \cite{Gu18} contained visual action labels where a person in a single video frame was labeled using the (visual-only) context of the surrounding 3-second interval. Since both datasets provide timestamps with labels, we can compute the co-occurrence of the labels in AVA v1.0 with the labels released here (AVA-Speech). Table \ref{tab:ava-cooccur-stats} demonstrates one such analysis, looking at the visual activities: ``talk to'', ``sing to'', ``listen-to-person'', ``listen (e.g., to music)'' and ``answer phone'' which can all be expected to correlate with audio speech. We can make a few observations. 

First, these visual label classes (except ``listen'') largely occur within an audio speech segment. However, the portion of occurrences that correspond to the NoSpeech audio label is significant, showing that visual and audio inferences by human labelers do not perfectly overlap. This lends further impetus to the idea that future approaches to activity detection designed to better understand scenes should consider audio-visual approaches. 

Second, the ``sing-to'' class from AVA v1.0 co-occurs most frequently with the SpeechWithMusic class in AVA-Speech, as one would expect, and many of the overlapping SpeechWithNoise segments also have music as part of the background. 

Finally, the ``listen'' and ``answer phone'' classes have high co-occurrences with NoSpeech. This matches intuition: ``listen'' class is specific to sounds other than human speech (speech is covered under the listen-to-person class), and phone answering begins before speaking into a phone. Here, we start seeing the temporal relationships between audio and visual actions.

\begin{table}[t]
    \centering
    \begin{tabular}{|l|c|c|c|c|}
        \hline         
            \hspace{0.2in} Label          & Time & Segments & AvgDur & SNR\\\hline
         CleanSpeech    & 14.55\% & 16.68\% & 2.97 sec & 40.8dB\\\hline
         Speech+Music  & 13.46\% & 13.33\% & 3.43 sec & 11.7dB\\\hline
         Speech+Noise  & 24.32\% & 25.41\% & 3.28 sec & 16.2dB\\\hline
         NoSpeech  & 47.68\% & 44.57\% & 3.68 sec & N/A\\\hline
    \end{tabular}
    \caption{Aggregate statistics over the AVA-Speech dataset.}
    \label{tab:data-stats}
\end{table}

\begin{table}[t]
    \centering
    \begin{tabular}{|l|r|r|r|r|}
        \hline         
         v1.0-Label    & Speech & +Music & +Noise & NoSpeech\\\hline
         talk-to & 22.7\% & 17.5\% & 36.9\% & 22.9\% \\\hline
         sing-to & 5.9\% & 56.6\% & 20.7\% & 16.8\% \\\hline
         listen-to-person  & 23.2\% & 15.0\% & 36.3\% & 25.5\% \\\hline
         listen & 0.0\% & 20.8\% & 19.1\% & 60.1\% \\\hline
         answer phone & 19.1\% & 16.6\% & 26.2\% & 38.2\% \\\hline
    \end{tabular}
    \caption{Co-occurrence percentages of speech-related AVA v1.0 activity labels (rows) with AVA-Speech speech segment labels (columns).}
    \label{tab:ava-cooccur-stats}
\vspace{-5mm}
\end{table}

\subsection{Speech-to-Noise Ratio Estimation}
To estimate speech-to-noise ratio (SNR), we apply a trained time-frequency-masking-based speech enhancement neural network, similar to the networks described in \cite{erdogan2015phase}, but consisting of stacked convolutional, bidirectional LSTM, and fully connected layers instead of only the bidirectional LSTM layers, trained on artificial mixtures of speech from LibriVox audio books and non-speech sounds from AudioSet \cite{gemmeke2017audio}. It provides two outputs, estimated speech and estimated non-speech, which we use to compute SNR as the ratio of the speech energy to the non-speech energy over the labeled segments.

The last column in Table \ref{tab:data-stats} shows the mean of the distribution of the speech-to-noise ratios across all segments for each label, based on the estimator. We note that the speech enhancement system is not a perfect model, and the SNR estimates should only be used as indicative of the relative difficulty of detecting speech across the different label types but cannot be treated as authoritative. It does indicate, as one would expect, that the distribution of the ``CleanSpeech" class is in a higher SNR range than the other speech classes, and the ``SpeechWithMusic" and ``SpeechWithNoise" overlap each other in a lower SNR range. As a result of the difference in the SNR ranges, we expect that the speech with overlapping music or noise is likely to be more difficult to detect than the CleanSpeech class. A few characteristics of the labeling process (and the guidance provided to the raters) result in the separation between CleanSpeech and SpeechWithMusic or Noise classes not being quite as clear as one would expect. Labelers were asked to identify all occurrences of speech activity, including hushed, low energy speech. Labelers don't always end speech segments when there are small gaps in the speech activity, and spot checks confirm presence of gaps in the speech labels, which lowers the predicted SNR for CleanSpeech segments. Finally, a number of the SpeechWithNoise segments consist of fairly low background noise, e.g., intermittent rustling or shuffling sounds (due to clothing/sheets or footsteps) are a common instance of background sounds for high SNR SpeechWithNoise segments.


\section{Benchmarking Speech Detection Models}
\label{sec:models}

In this section, we benchmark off-the-shelf speech detection models based on both audio-only and visual-only inputs, without any fine-tuning improvements for the AVA-Speech dataset.

\subsection{Acoustic models}

We use the voice activity detector of WebRTC \cite{WebRTC} system as a publicly-available baseline (RTC\_vad).  We also report results for a state-of-the-art acoustic speech detector using convolutional neural networks (CNNs) \cite{hershey2017cnn} trained on AudioSet data \cite{gemmeke2017audio}, over 1M 10-second excerpts from YouTube videos manually labeled for speech presence. We report results from two versions:  tiny\_320 has 3 convolutional layers, $<$1M weights, uses 32 frames of a 10ms-per-frame 64-band mel spectrogram as input, computes 23M multiplies per inference.  resnet\_960 use the larger ResNet-50 architecture, uses 96 frames of input, consists of 30M weights, and computes 1900M multiplies per inference. The AudioSet training data includes over 500 classes, and the models were trained to optimize over the entire set. For this evaluation, we only used the speech class output.

\subsection{Visual Speech Classification models}
We perform visual speech classification (VSC) by first detecting and tracking faces in the video, followed by applying a stacked CNN model on every set of 3 consecutive face thumbnails from each track to classify speech/non-speech. The CNN model architecture, a sequence of depthwise convolutional layers followed by an average pooling layer and fully connected layers and a softmax classification layer for a total of 46K parameters, is motivated by MobileNet \cite{howard2017mobilenet}, but uses the same number of filters for each depthwise convolutional layer since activity recognition needs higher capacity in the earlier stages to capture motion. VSC models predict whether each face is speaking; however, as the AVA-Speech dataset doesn't attribute speaking information to faces, we aggregate the model scores across all the visible faces at each instant and keep the max score as the model prediction for Speech-Active. 

\subsection{Model Performance}
We report performance on our dataset as a baseline for future reference.  Since the WebRTC VAD has a false-positive rate (FPR) of 0.315 on our NoSpeech segments, we report true-positive rates (TPR) for all models tuned to the same FPR.  We break out the evaluation in 2 ways: (1) For each speech condition, $\{$CleanSpeech, SpeechWithMusic, SpeechWithNoise$\}$, we calculate TPR at the frame level contrasted with $NoSpeech$ as the negative class. (2) We combine all the frames for the 3 speech conditions into a single positive class (``Speech''). Table \ref{tab:model-perf-table} shows the performance of the different models across the 4 conditions using TPR. As we expect, the performance of audio models degrades as background noise levels increase, while the performance of VSC remains relatively constant.

The last column of Table \ref{tab:model-perf-table} reports benchmarks for comparing average inference latency. The VSC model latency includes the expensive face detection and tracking time in addition to model inference. All results are obtained on a workstation with a 2.60GHz Intel Xeon E5-2690 CPU (turbo boost off), 128GB RAM; inferences are performed in single thread mode. 

\begin{figure}[t]
  \centering
  \includegraphics[width=\linewidth]{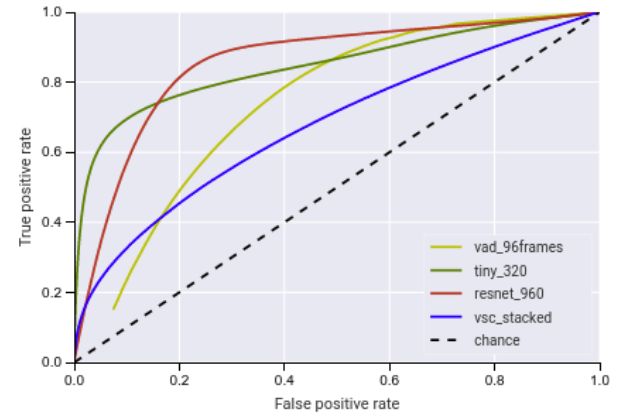}
  \caption{ROC curves for several speech detection models, for Speech vs NoSpeech classification.}
  \label{fig:results_roc}
\end{figure}

We recognize that the choice of operating points for speech detection is heavily dependent on the downstream applications. To provide a view of performance across the space of possible operating points, Figure \ref{fig:results_roc} shows the ROC curve for each model, obtained by varying the classification threshold for distinguishing NoSpeech from the Speech classes. Note that while the VSC curve is below those pf the audio models, the current evaluation framework is inherently disadvantageous to the VSC model, since it only considers audible speech. As shown in Table \ref{tab:ava-cooccur-stats}, human labels in AVA v1.0 for talk/sing-to events only co-occur with audio speech labels about $77\%$ of the time, and the performance of the VSC model should be considered in that context rather than as a direct comparison to the audio models.

\begin{table}[]
    \centering
    \footnotesize
    \begin{tabular}{|l|c|c c c c|c|}
        \hline
         & FPR & \multicolumn{4}{c|}{TPR} & Time \\
         Model  &  & Clean & Noise & Music & All &  (ms) \\\hline
         RTC\_vad    & 0.315 & 0.786 & 0.706 & 0.733 & 0.722 & 0.06 \\
         tiny\_320   & 0.315 & 0.965 & 0.826 & 0.623 & 0.810 & 2.23\\
         resnet\_960  & 0.315 & 0.992 & 0.944 & 0.787 & 0.917 & 265 \\\hline
         VSC  & 0.315 & 0.588 & 0.568 & 0.556 & 0.569 &  40.0\\\hline
    \end{tabular}
    \caption{False and True Positive Rates for the different models.}
    \label{tab:model-perf-table}
\end{table}

\section{Conclusions}
\label{sec:conclusion}

This paper introduces AVA-Speech, a densely and temporally labeled speech activity dataset with background noise conditions annotated for about 46 hours of movie video, covering a wide variety of conditions containing spoken activity, that we will release publicly as part of the AVA website. We presented performance baselines on this dataset using off-the-shelf, state-of-the-art approaches to audio-based and image-based speech activity detection models. We anticipate that this video dataset will spur further interesting research in joint audio and visual models, and hope to contribute further to the development of additional labeling and models on this dataset, to enable its use as a shared standard for tasks beyond speech activity detection. AVA 2.0 is under construction and we will augment this dataset with speech activity labels on those videos as well. 

\bibliographystyle{IEEEtran}
\bibliography{is2018bib}

\end{document}